\documentclass[a4paper,12pt,reqno,superscriptaddress]{revtex4}

\usepackage{graphicx}
\usepackage{dcolumn}
\usepackage{epsfig}

\usepackage[centertags]{amsmath}
\usepackage{amsfonts}
\usepackage{euscript}
\usepackage{amssymb}
\usepackage{amsthm}
\usepackage{newlfont}
\usepackage{stmaryrd}
\usepackage{mathrsfs}
\usepackage{subfigure}

\let\oldmarginpar\marginpar
\renewcommand\marginpar[1]{\-\oldmarginpar[\raggedleft\footnotesize #1]%
{\raggedright\footnotesize #1}}

\newcommand{\ep}{\varepsilon}

\let\al=\alpha \let\be=\beta  \let\ep=\epsilon
\let\ve=\varepsilon   
\let\ka=\kappa \let\la=\lambda  
  \let\th=\theta

\let\De=\Delta \let\Ga=\Gamma

\newcommand{\opunit}{\text{1}\kern-0.22em\text{l}}

\DeclareMathAlphabet{\mathpzc}{OT1}{pzc}{m}{it}

\newcommand{\cf}{\textit{cf.}}

\newcommand{\id}{\textrm{d}}

\begin{document}

\title{Heat capacity in nonequilibrium steady states}

\author{E.~Boksenbojm}
\email{eliran.boksenbojm@fys.kuleuven.be}
\affiliation{Instituut voor Theoretische Fysica, K.U.Leuven, Belgium}
\author{C.~Maes}
\affiliation{Instituut voor Theoretische Fysica, K.U.Leuven, Belgium}
\author{K.~Neto\v{c}n\'{y}}
\affiliation{Institute of Physics AS CR, Prague, Czech Republic}
\author{J.~Pe\v{s}ek}
\affiliation{Institute of Physics AS CR, Prague, Czech Republic}
\affiliation{Faculty of Mathematics and Physics, Charles University in Prague, Czech Republic}




\begin{abstract}
We show how to extend the concept of heat capacity to nonequilibrium systems. The main idea is to consider the excess heat released by an already dissipative system when slowly changing the environment temperature.  We take the framework of Markov jump processes to embed the specific physics of small driven systems and we demonstrate that heat capacities can be consistently defined in the quasistatic limit.
Away from thermal equilibrium, an additional term appears to the usual energy--temperature response at constant volume, explicitly in terms of the excess work. In  linear order around an equilibrium dynamics that extra term is an energy--driving response and it is entirely determined from  local detailed balance. Examples illustrate how the steady heat capacity can become negative when far from equilibrium.
\end{abstract}

\maketitle

\newpage
\section{Introduction}

The study of thermophysical properties of materials has played a major role in the development of thermodynamics and physics in general. A key issue is to understand how the system responds to variations in external control fields via heat exchange with its environment. The discussion simplifies for reversible thermal processes that are slow enough and pass through a sequence of equilibrium states. The heat exchange along such a process is determined by the way the system accommodates to the modified external conditions and relaxes to the new equilibrium state. Restricting to processes parameterized by temperature while other parameters (like volume, pressure etc.) are fixed, leads to the notion of heat capacity as a primary quantifier of the heat exchange. Their determination and characterization
has proven very relevant in a great variety of domains ranging from industrial applications, over the study of phase transitions to fundamental tests for understanding the relation between mechanics and thermodynamics.
Not surprisingly they were also key objects of study in the beginnings of quantum theory, in the further development of solid state physics and in the thermodynamics of new materials.

There is no well-established nonequilibrium theory.
So far, the study of nonequilibrium heat capacities and related quantities has been mostly restricted to transient systems.  There, internal relaxation is slow compared to the time-dependent control fields, with glassy systems as a paradigmatic example.  A standard approach to transient systems involves frequency-dependent heat capacities as
analyzed in several theoretical and experimental studies~\cite{gard,sta}.
In contrast, the present letter considers systems that are well relaxed but under stationary nonequilibrium conditions. The study of heat capacities for such systems is largely unexplored. Such studies would include the thermal conditions of active matter; it would ask for heat capacities of bodies in which life processes take place, and
it would seem to require nonequilibrium extensions of thermodynamic potentials.  These questions are probably too difficult and too broad to answer at
once, but nevertheless they motivate us in the initial set-up and in our modeling.  A central issue is then whether and how the steady
nonequilibrium functioning produces substantial deviations from the equilibrium heat capacity, not because the system has not fully relaxed but
because of a totally different physics altogether.\\

Preliminary calculations of steady state heat capacities within the framework of linear irreversible thermodynamics to explain certain involved conduction calorimetry experiments have been reported since almost twenty years~\cite{cera}.  We follow here more closely the ideas that were developed more recently in~\cite{seki,house,stt,ha}, which is sometimes referred to as steady state thermodynamics.

To be specific, we mostly stick to a discrete set-up with a
driven stochastic dynamics that is consistent with the presumed microscopic reversibility via the principle of \emph{local} detailed balance and which
covers a wide range of physically relevant nonequilibrium processes.
As an example, we discuss at the end a model of driven diffusion in one and two dimensions, that naturally fits our formalism via the continuous (diffusion) limit of discrete approximations.\\

\section{Nonequilibrium model}

We have in mind small thermodynamically-open systems on which mechanical work is performed and which are coupled to an environment represented by a single heat bath. A crucial physical hypothesis is that the external forces are not fully conservative so that the system is always dissipative. Our aim is to analyze to what extent the heat exchanged with the bath while slowly changing its temperature can be represented by a well defined heat capacity.

To be specific, we consider Markovian dynamics with discrete states $x, y,\ldots$ representing distinct (mesoscopic) configurations of the system.
It is a stochastic process with trajectories over a time-interval $[0,\tau]$ written as
\begin{equation}\label{eq: trajectory}
  [x_t] = (x^0 \stackrel{t^1}{\rightarrow} x^1 \stackrel{t^2}{\rightarrow} \ldots \stackrel{t^n}{\rightarrow} x^n)\,,\quad
  0 < t^1 < \ldots < t^n < \tau
\end{equation}
each specified by a sequence of random jumps between states.
Each state $x$ is given an energy $E(x)$ representing
all conservative forces acting on the system. The non-conservative forces
need to be introduced via the amount of work $F(x,y) = -F(y,x)$ they perform on the system when it jumps from state $x$ to $y$.
Here we mostly assume that the energy function $E(x)$ and the non-conservative work function $F(x,y)$ are constant in time and we concentrate on the thermodynamic process corresponding to (slow) changes in the bath temperature $T(t)$. (For $F=0$ this would lead to the heat capacity at constant volume.)

Energy conservation on the level of a single trajectory $[x_t]$, $\,0 \leq t \leq \tau,$ can be written as
\begin{equation}\label{first-law}
  E(x_\tau) - E(x_0) = W_F([x_t]) + Q([x_t])
\end{equation}
Here the change of energy is decomposed into the work of the non-conservative forces, $W_F$, and the heat $Q$ flowing into
the system. The work of conservative forces is zero by our assumption that $E(x)$ does not explicitly depend on time.
The work of non-conservative forces is
\begin{equation}
  W_F([x_t]) = \sum_{j=1}^n F(x^{j-1},x^j)
\end{equation}
with the sum over all jump times in the trajectory~\eqref{eq: trajectory}.
Since $F$ does not derive from a potential, the work $W_F$ remains a non-trivial trajectory-dependent function which is $\tau-$extensive for typical
paths; the
same being true for the heat $Q$ for which the balance relation~\eqref{first-law} serves as a definition.\\

The dynamics is determined by transition rates $\la^\be(x,y)$ that are time-dependent through their explicit dependence on the inverse temperature $\beta = 1/T$ (setting the Boltzmann constant to unity),
\begin{equation}\label{psip}
  \la^\be(x,y) = \psi^\be(x,y)\,
  e^{\frac{\be}{2}\,[E(x) - E(y) + F(x,y)]}
\end{equation}
By the condition of local detailed balance (expressing thermal equilibrium in the
coupled heat bath, see~\cite{KLS}),
the symmetry condition $\psi^\be(x,y) = \psi^\be (y,x)$ has to be always satisfied.
If we now vary the temperature in time, the time-dependent distribution $\rho_t(x)$ solves the Master equation
\begin{equation} \label{mastereq}
  \dot\rho_t(x) =
  \sum_{y} [\rho_t(y) \lambda^{\be(t)}(y,x) - \rho_t(x) \lambda^{\be(t)}(x,y)]
\end{equation}
We assume that for a fixed inverse temperature $\be$ the stationary distribution $\bar\rho^\be$ is unique and approached exponentially fast with relaxation time $\tau_R$;  the latter provides a reference time-scale to delineate the quasistatic regime. Expectations with respect to $\bar\rho^\be$ will be denoted by
$\langle \cdot \rangle^\be$.

An essential feature of our model is that its stationary regime is fundamentally different from equilibrium: despite the local detailed balance, one has
$\bar\rho^\be(x)\,\la^\be(x,y) \neq \bar\rho^\be(y)\,\la^\be(y,x)$ unless $F$ derives from a potential. In particular, the system exhibits steady dissipation, the rate of which is given by the (positive) mean stationary work (or equally heat) per unit time $\langle w^\beta \rangle^\be$, the expectation value of $w^{\beta}(x) = \sum_y \la^\be(x,y)\,F(x,y)$ which is the expected power of the non-conservative forces when the system is in state $x$. Note that we allow the transition rates to depend on time only via their temperature-dependence --- this condition will simplify the construction of the quasistatic limit. \\

\section{Steady heat capacity}

We come to our main question:
Under what conditions and in what sense can \emph{some} averaged heat $\langle Q \rangle$ along a process corresponding to $T(t)$, $0 \leq t \leq \tau$, be
given the form $\int C_F\,\id T$, with $C_F$ an appropriate heat capacity? This can only be true provided that $\langle Q \rangle$ is
`geometric' in the sense that it only depends on the values of temperature and not on how fast $T$ changes in time. Such a property is known to hold for currents in the quasistatic limit of infinitely slow process~\cite{pump}, irrespective of being in or out of equilibrium. However, the essential difference between the equilibrium and the nonequilibrium cases is that in the latter there are non-zero stationary (sometimes called `house-keeping') currents which are due to the intrinsic dissipation of the nonequilibrium steady states. 
These non-geometric currents need to be regularized away to
separate the \emph{excess} currents that are to be seen as a natural extension of the equilibrium energy changes.

Next we explain in detail how this can be applied to construct the steady heat capacity in a consistent way. We more closely follow the formalism of
Ref.~\cite{ha} using the terms `house-keeping heat' and the `excess heat' for the 
stationary and the geometric components, respectively.\\

On a somewhat intuitive level, the thermodynamic process induced by changing
$\be(t)$, $0 \leq t \leq \tau$, can be considered to be
quasistatic provided that the whole time interval can be suitably discretized,
$\tau = N \Delta\tau$, so that (1) $\De\tau \gg \tau_R$
(relaxation time), and (2) $|\Delta \be| /\be \ll 1$
over all elementary time-intervals.
Whenever $\tau \gg \tau_R$, such a discretization is possible and we can see the whole process as essentially consisting
of a sequence of $N$ sudden and small temperature changes $\De\be = O(1/N)$, each one followed by relaxation to the new steady conditions.

Within the $k-$th time interval $[\tau_{k-1}, \tau_k] = [k-1,k]\,\De \tau$, the
system can be thought to relax to the steady state distribution $\bar\rho^{\be(\tau_k)}$, starting at time $\tau_{k-1}$ from the steady distribution $\bar\rho^{\be(\tau_{k-1})}$ reached in the previous interval. Up to leading order, the initial distribution can be given in terms of the final steady distribution as
\begin{equation}\label{qs}
  \bar\rho^{\be(\tau_{k-1})} = \bar\rho^{\be(\tau_k)} -
  \frac{\partial\bar\rho^\be}{\partial \be}\,[\be(\tau_k) - \be(\tau_{k-1})] +
  O(N^{-2})
\end{equation}
while the expected work of the non-conservative forces within the relaxation process equals
\begin{equation}\label{kw}
  \Delta_k W_F =   \int_{\tau_{k-1}}^{\tau_k}
  \sum_x \rho_t(x)\,w^{\be(t)}(x)\,\id t
\end{equation}
where
$w^{\be(t)}(x) = \sum_y \la^{\be(t)}(x,y)\,F(x,y)$ is the expected power at time $t$
provided the system is in state $x$.
Approximating $w^{\be(t)}$ within the entire interval $[\tau_{k-1},\tau_k]$ by $w^{\be(\tau_k)}$, we rewrite~\eqref{kw} up to corrections $O(N^{-2})$ in the form
\begin{equation}
 \Delta_k W_F =
  \Delta \tau \sum_x \bar\rho^{\be(\tau_k)}(x) \,w^{\be(\tau_k)}(x)
  + \int_{\tau_{k-1}}^{\tau_k} \sum_x \left[ \rho_t(x) - \bar\rho^{\be(\tau_k)}(x)\right] \,w^{\be(\tau_k)}(x)\,\id t
\end{equation}
The first term on the right-hand side of this equation is the house-keeping part of the work. It corresponds to the expected work if the system would
be in its stationary state at every instant of time. The other term corresponds to the excess work. By assumption, $\De\tau \gg \tau_R$ and hence the system does reach the stationary state $\bar\rho^{\be(t_{k})}$, which means that we can as well take the upper limit of the integral to be $+\infty$.
Formally solving the Master equation~\eqref{mastereq} to find $\rho_t$ and after some standard manipulation, we obtain the total quasistatic work of non-conservative forces by summing over $k$:
\begin{equation}\label{work-quasistatic}
  \langle W_F \rangle = \int_0^\tau \bigl\langle w^{\be(t)} \bigr\rangle^{\be(t)}\,\id t
  +\int \Bigl\langle \frac{\partial}{\partial\be}\, V^{\be} \Bigr\rangle^\be\, \id\be
  + O\bigl( \frac{\tau_R}{\tau} \bigr)
\end{equation}
with
\begin{equation}
  \langle w^\beta\rangle^\beta = \frac{1}{2} \sum_{x,y}
  F(x,y)\,[\bar\rho^\be(x)\,\la^\be(x,y) - \bar\rho^\be(y)\,\la^\be(y,x)]
\end{equation}
the steady rate of dissipation, given in the standard `force times current' form.  The first term in~\eqref{work-quasistatic} is therefore the steady state (or `house-keeping') component.  The second term on the other hand relates to the transient
(or `excess') component where we have introduced
\begin{equation}\label{work-relaxation}
  V^{\be}(x) = \int_0^\infty [\langle w^\be(x_t)\rangle_{x_0=x}
  - \langle w^{\be} \rangle^\beta ]\,\id t
\end{equation}
The state function $V^{\be}(x)$ is to
be understood as the transient part of the mean dissipated work along the complete relaxation path started from state $x$.  The function $\langle w^\be(x_t)\rangle_{x_0=x}$ yields the expected power at time $t$ given that the system was started in state $x$ at time zero. Note that
$\langle w^\be(x_t)\rangle_{x_0=x} \simeq \langle w^{\be} \rangle^\beta$
for times $t\gg \tau_R$, and $\langle V^\be \rangle^\be = 0$.

In the same quasistatic regime where the system essentially passes through a succession of steady states, the expected change in energy is
$\langle E(x_\tau) - E(x_0) \rangle = \int \frac{\partial}{\partial \be}\,
\langle E \rangle^\be\,\id \be$. Hence, from the First Law~\eqref{first-law},
$\langle Q \rangle = -\int_0^\tau \langle w^{\be(t)} \rangle^{\be(t)}\,\id t
+ \langle Q \rangle^{ex} + O(\tau_R / \tau)$ with the excess heat
$\langle Q \rangle^{ex} = \int C_F\,\id (1/\be)$ given in terms of the generalized heat capacity
\begin{equation}\label{heat-capacity}
  C_F = -\be^2\frac{\partial}{\partial\be}\,\langle E \rangle^\be
  +\be^2\Bigl\langle \frac{\partial}{\partial \be} \,V^{\be} \Bigr\rangle^\beta
\end{equation}
This is our main result.
The first term resembles the familiar equilibrium expression for the heat capacity at constant volume (and/or other external parameters) but now under the nonequilibrium steady state.
The second term is novel and it originates from the fact that even keeping all the external parameters and forces fixed and merely changing
the temperature, there is an extra non-zero work done. Part of the energy which is added to the system can be used to change the stationary currents, reminiscent of the more familiar Mayer relation between the heat capacities at constant volume and pressure. In general, the function $V^\be(x)$ non-trivially couples both variables $\be$ and $x$ so that, without further conditions, the heat capacity cannot be written as the temperature derivative of some generalized thermodynamic potential such as in the construction of (equilibrium) enthalpy.

The above derivation of formula~\eqref{work-quasistatic} can easily be turned into a rigorous argument~\cite{kaji}: Using the
quasistatic (or `adiabatic') scaling of the time-dependent protocol,
$T(t) \mapsto T(\ve t)$, $\tau \mapsto \ve^{-1}\tau$, both the work of
non-conservative forces and the heat can be systematically expanded in powers of $\ve$. In this framework the house-keeping part is recognized as
a linearly diverging term of order $\ve^{-1}$ and the non-quasistatic corrections are $O(\ve)$. The excess work/heat are the finite
(or `renormalized') parts of both by construction diverging quantities. It is precisely in this sense that they can be considered as well-defined.\\

\section{Experimental access}

It is crucial for the consistency of our construction that $C_F$ is defined through
\emph{excess} heat that was proven to be geometric, i.e., fully determined by the steady state properties. In principle, both the steady rate of
dissipation $\langle w^\be \rangle^\beta$ and the transient work functions $V^\be(x)$ along relaxation paths can be obtained by measurement,
independently of measuring the heat capacity from the quasistatic heat exchange.
In this way, a specific prediction is given concerning the mutual relation between the
results of \emph{a priori} different types of experiment.

Clearly, the experimental accessibility of the generalized capacity strongly depends on whether the excess heat in the decomposition
$\langle Q \rangle = -\int_0^\tau \langle w^{\be(t)} \rangle^{\be(t)}\,\id t
+ \langle Q \rangle^{ex} + O(\tau_R / \tau)$
can be distinguished from the house-keeping (diverging) component
$-\int_0^\tau \langle w^{\be(t)} \rangle^{\be(t)}\,\id t$. A natural possibility comes from the different symmetry
properties of the contributions: under the protocol reversal
$\be(t) \mapsto \be(\tau - t)$, the house-keeping part is symmetric whereas the excess
part is antisymmetric; the residual non-quasistatic corrections have no definite protocol-reversal behavior. Hence, the excess heat along any path
can in principle be extracted by repeatedly traveling the same temperature-path back and forth and counting in only differences in the heat exchanged or the work done. At the same time, the temperature changes need to be slow enough to avoid non-quasistatic residuals.
Estimating the experimental errors with such a procedure probably remains a challenging but very relevant and physically interesting problem.\\

Naturally, the same questions as discussed here in the context of stochastic systems can be addressed for macroscopic bodies under nonequilibrium conditions. A particular experimental setup has already been proposed in Ref.~\cite{cera}: The authors there employ conduction calorimetry techniques to study the heat produced by a ferroelectric sample heated by an applied high-frequency AC-current. Changing the environment temperature results in modifications of the outgoing heat current that can be directly measured in real time by an imposed thermopile. It is argued that the method is subtle enough to distinguish the steady heat currents from their excess components which, after some time-integration, yield the heat capacity by definition. In this context, the present letter provides a general microscopic (or, more precisely, mesoscopic) theory for such a type of experiments on dissipative systems.\\

\section{Linear nonequilibrium correction}

Some progress can be made in a close-to-equilibrium regime where we can control the steady-state properties of the system by a systematic expansion
in the magnitude of the nonequilibrium driving. In the pioneering work of McLennan~\cite{mac}, he found leading nonequilibrium corrections to the canonical
distribution in terms of entropy changes; see also~\cite{KN,mat} for recent extensions. Here we use the formulation and results of Ref.~\cite{jmp}.

For small non-conservative forces $F$, the stationary distribution
$\bar\rho^\be$ can be well approximated by the McLennan ensemble,
\begin{equation}\label{mclennan}
  \bar\rho^\be(x) \simeq \frac{1}{Z^\be}\,\exp\,[-\be E(x) - \be V^{\be}(x)]
\end{equation}
in which the correction term exactly coincides with the dissipated work along relaxation paths~\eqref{work-relaxation}. This formula can be justified by
scaling the driving forces as $F(x,y) \mapsto \ep F(x,y)$ and expanding in powers of $\ep$; the McLennan distribution is proven correct up to order
$\ep$.

Up to linear order in $F$ and by the construction
of $V^\be$, the nonequilibrium term in~\eqref{heat-capacity} can be written in the form of equilibrium time-correlations between the energy and the power of the
non-conservative forces :
\begin{equation}
  C_F \simeq -\be^2\frac{\partial}{\partial\be}\,\langle E \rangle^\be -
  \be^2 \int_0^\infty \langle E(x_0)\,w^\be(x_t) \rangle^\beta_{\text{eq}}\,\id t
\end{equation}
Here the expectation $\langle \cdot \rangle^\be_{\text{eq}}$ is under the equilibrium distribution $\bar\rho^\be_{\text{eq}}(x) = \exp\,[-\be E(x)] / Z^\be$ and we have used that $\langle w^\be \rangle^\beta_{\text{eq}} = 0$.
Finally, combining with the McLennan formula~\eqref{mclennan}, we finally obtain the relation
\begin{equation}\label{thre}
  C_F \simeq -\be^2 \frac{\partial}{\partial\be}\,\langle E \rangle^\be -
  \be\,(\langle E \rangle^{\be} - \langle E \rangle^{\be}_{\text{eq}})
\end{equation}
always correct up to linear order in the nonequilibrium driving.
Hence the close-to-equilibrium heat capacity consists of two linear-response contributions: (1) the
(equilibrium-like) energy--temperature response and (2) the  energy--driving response, which can be further rewritten in terms of an equilibrium correlation function like in the Green-Kubo relation.
While the quasistatic heat capacity on the left-hand side of \eqref{thre} derives from a thermodynamic process, the two response-functions on the right-hand side are by definition steady-state properties of the system. All three quantities in~\eqref{thre} are independently measurable, at least \emph{in principle}.
Note there is no dependence on the symmetric part $\psi^\be$ in the transition rates~\eqref{psip}, which is at the origin of the remarkable simplification in the close-to-equilibrium regime.

Remark that this linear order theory is only meaningful when the dynamics breaks the driving-reversal symmetry $F \mapsto -F$ (simultaneously for all transitions $x \leftrightarrow y$). Under this symmetry, the linear nonequilibrium corrections to the heat capacity vanish,
$C_{\ep F} = C_0 + O(\ep^2)$, due to the absence of the $O(\ep)$ corrections in both the mean energy $\langle E \rangle^\be$ and the transient work function $V^\be$. The higher-order corrections can be obtained by a systematic expansion in powers of the parameter $\ep$ adopted to control the distance from equilibrium~\cite{mat,kaji}.\\

The general non-perturbative formula~\eqref{heat-capacity} and its close-to-equilibrium approximation~\eqref{thre} for driving-reversal asymmetric systems constitute the main results of this letter. In the next section we give specific examples that go beyond the scope of the simple linear theory and on which we demonstrate some peculiar features of the steady heat capacity~\eqref{heat-capacity}.\\

\section{Example: driven diffusion}

As a trial nonequilibrium system we consider the case of independent colloids driven in a toroidal trap, which is experimentally feasible~\cite{cil}. The particle motion can be modeled by the overdamped driven diffusion on a circle of unit length,
\begin{equation}\label{1d}
  \dot{x}_t = F - E'(x_t) + \sqrt{2T(t)}\,\xi_t
\end{equation}
($\xi_t$ is standard white noise.) The driving force $F$ is constant and, to be specific, we take the potential landscape $E(x) = \sin (2\pi x)$.
\begin{figure}[t]
  \includegraphics[width=.8\textwidth,height=!]{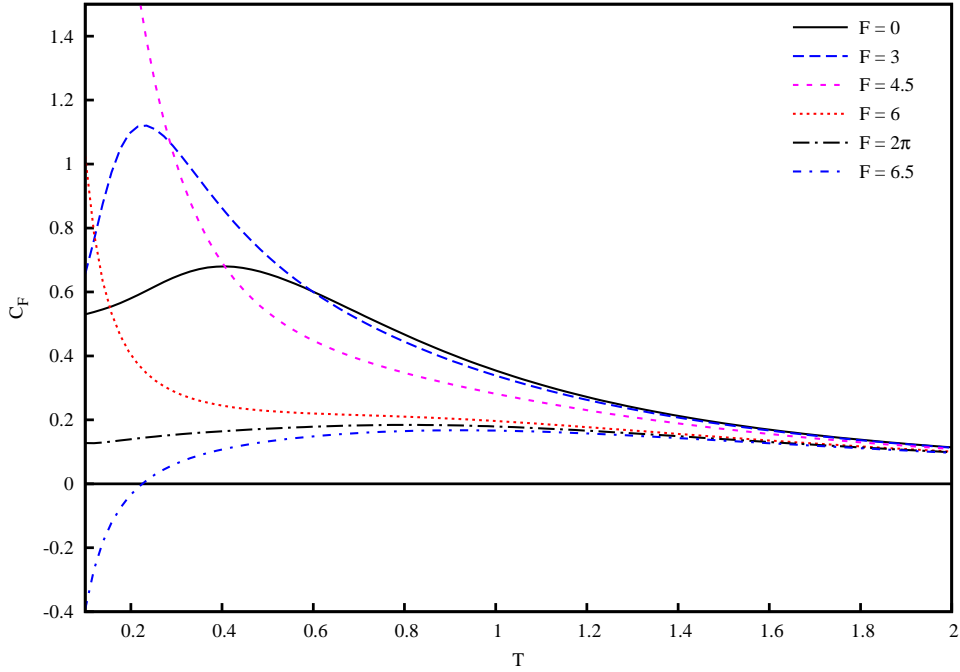}
  \caption{Steady heat capacity $C_F$ for driven one-dimensional diffusion. The trial potential landscape is $E(x) = \sin(2\pi x)$. }
\label{fig: CF}
\end{figure}
The steady heat capacity $C_F$ is depicted in Fig.~\ref{fig: CF}; we have evaluated~\eqref{heat-capacity} numerically exactly for a discrete-space approximation of the dynamics~\eqref{1d}. For large temperatures the nonequilibrium correction to the steady heat capacity becomes dominated by the energy-temperature response (the first term in~\eqref{heat-capacity}) and $C_F(T)$ asymptotically approaches the equilibrium curve
$C_0(T) = 1 /(2 T^2) + O(T^{-3})$, for arbitrary forcing $F$. On the other hand, at lower temperatures the nonequilibrium correction becomes relevant and we see a qualitative change of behavior across the value $F^* = 2\pi$, which we associate with the crossover between the limiting fixed point and the limiting cycle in the zero-temperature (deterministic) solution of~\eqref{1d}.

Our model demonstrates that $C_F$ can obtain negative values when far from equilibrium. Although similar observations concerning negative heat capacities have been made before for systems non-weakly coupled to finite reservoirs~\cite{hanggi}, here the physical origin is different and the effect emerges due to the nonequilibrium nature of our system; see more below.

\begin{figure}[t]
  \includegraphics[width=.8\textwidth,height=!]{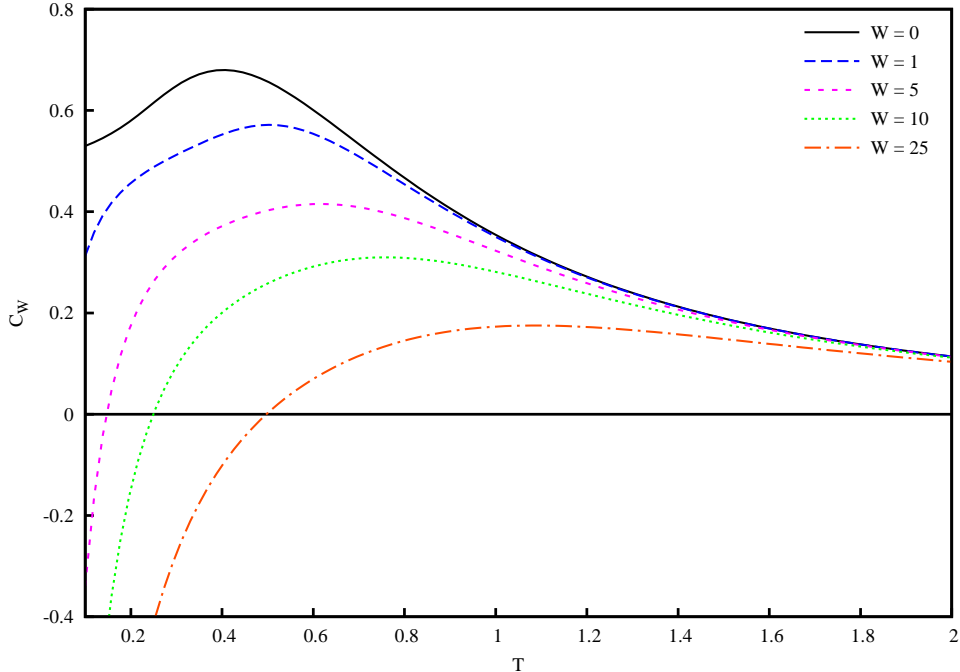}
  \caption{Steady heat capacity $C_W$ for the one-dimensional diffusion model.}
  \label{fig: CW}
\end{figure}

We also calculate the steady heat capacity at constant steady power, $C_W$, as defined via the excess heat along the quasistatic curve $(T,F)$ on which the steady power $\langle w^\beta \rangle^\be$ remains constant. The general relation between both heat capacities is readily found to be
\begin{equation}\label{mayer}
  C_W = C_F
  + \be^2 \frac{\partial\langle w^\beta \rangle^\be}{\partial \be}\,
  \Bigl( \frac{\partial\langle w^\beta \rangle^\be}{\partial F} \Bigr)^{-1}
  \Bigl[ \frac{\partial \langle U \rangle^\be}{\partial F} -
  \Bigl\langle \frac{\partial}{\partial F} \,V^{\be} \Bigr\rangle^\be
  \Bigr]
\end{equation}
with $W$ and $F$ related by the condition $\langle w^\beta \rangle^\be = W$. For our diffusion model~\eqref{1d}, the heat capacity $C_W$ as a function of temperature is depicted in Fig.~\ref{fig: CW}.\\

To get a better understanding of how the steady heat capacity depends on the dissipative properties of the system, we further consider the two-dimensional modification of the model~\eqref{1d},
\begin{equation}
  \dot{X}_t = \vec F(X_t) - \nabla E(X_t) + \sqrt{2T(t)}\,\vec\xi_t\,,\quad
  X = (x,y)
\end{equation}
with the spherically symmetric potential $E(X) = \frac{\la}{2}\,r^2$, $\la > 0$ and driven by the purely rotational field $\vec F(X) = \ka\,r^\al\,\vec e_\th$ with some $\al > -1$; the standard polar coordinates $r$ and $\th$ being used here. The conservative and the non-conservative fields are mutually orthogonal, $\vec F \cdot \nabla E = 0$, and the stationary density is insensitive to the nonequilibrium driving, \
$\bar\rho^\be = \exp(-\be E)/Z^\be$, i.e., the same as if the system were in equilibrium. Hence, the first term in the steady heat capacity~\eqref{heat-capacity} equals unity by equipartition. However, different steady states are distinguished by their mean dissipative power that equals
$\langle w^\beta \rangle^\be = \Ga(\al + 1)\,\ka^2\, (2 T / \la)^\al$.
The nonequilibrium correction term in $C_F$ can also be calculated analytically to yield the formula
\begin{equation}
  C_F = 1 + \frac{1}{2\la}\,\frac{\partial \langle w^\beta \rangle^\be}{\partial T}
\end{equation}
This simple relation between the steady heat capacity and the mean power is not to be expected in general.
Nevertheless, the relation makes it very clear that the steady heat capacity depends on how the dissipation, and not just the energy, depends on temperature.
In our model the increase of temperature makes the steady states less localized around the origin and depending on whether $\al > 0$ or $-1 < \al < 0$, this corresponds to a higher, respectively lower amount of dissipation as quantified by the mean power
$\langle w^\beta \rangle^\be$.  As a result, the nonequilibrium correction to the heat capacity obtains the same positive, respectively negative sign. This suggests that negative steady heat capacities may generally emerge for far-from-equilibrium systems when their steady dissipation decreases sufficiently strongly with temperature --- details are left to further studies. We conclude our short analysis of this model by noting the equality $C_W = C_F$ due to the $F-$independence of the stationary density $\bar\rho^\be$, \cf~formula~\eqref{mayer}.\\

\section{Conclusion and open questions}

We have analyzed a meaningful and consistent generalization of heat capacity to nonequilibrium systems. By applying and adapting the
previously developed framework of slow transformations of nonequilibrium steady states, we have derived the basic properties of the heat capacity
defined from the quasistatic heat. This construction makes physical sense because the finite excess part of the heat exchange is well-defined and geometric.  In formula~\eqref{heat-capacity} a general non-perturbative expression for the
steady heat capacity is given in terms of the (standard) energy--temperature response but modified with a new correction intimately related to the relaxation properties of the dissipative effects --- the new term derives from the transient work of the driving forces along relaxation paths.

We have demonstrated via simple examples that the steady heat capacity can take negative values as well. It has been argued that this phenomenon has to do with a specific temperature-dependence of dissipative characteristics far from equilibrium. The details of this proposal need to be further analyzed. Another relevant question is a detailed analysis of the steady heat capacity at low temperatures, in particular in regimes where the reference zero-temperature dynamical system is fundamentally different from the one in equilibrium. We expect the steady heat capacity and related nonequilibrium response functions to reveal important information about the presence of nonequilibrium phase transitions in the system~\cite{marro}.

We have also found more specific expressions for the heat capacity of close-to-equilibrium systems breaking the driving-reversal symmetry. In that case the nonequilibrium contribution to the heat capacity is directly related to the equilibrium linear response to switching on a (weak) nonequilibrium driving, see formula~\eqref{thre}. Equivalently, it can be given in terms
of equilibrium time-correlations resembling the Green-Kubo or fluctuation-dissipation relations.

To conclude, remark that presently the inertial degrees of freedom (the particles' momenta) have been considered `fast' with respect to `slow' spatial configurations, in the usual sense of time-scale separation. By this assumption, the distribution of momenta is always Maxwellian and the contribution to the total steady heat capacity follows the equipartition theorem as  $k_B /2$ per  momentum degree of freedom, in the exact same way as in equilibrium. This restriction is not essential and the momenta degrees of freedom with more general stationary distributions can easily be included in the theory.


\begin{acknowledgements}
We are grateful to Bram Wynants for initial work. E.B. and C.M. acknowledge
financial support from the FWO project G.0422.09N.
K.N.~acknowledges the support from the Grant Agency of the Czech Republic,  Grant no.~202/08/0361. J.P.~benefits from the Grant no.~51410 (the Grant Agency of Charles University) and from the project SVV-263301 (Charles University).
\end{acknowledgements}


\end{document}